\documentclass[a4paper,10pt,twoside]{cpc-hepnp}

\usepackage{multicol}
\usepackage{graphicx}
\usepackage{booktabs}
\usepackage{amssymb,bm,mathrsfs,bbm,amscd}
\usepackage[tbtags]{amsmath}
\usepackage{lastpage}

\begin{document}

\fancyhead[c]{}

\footnotetext[0]{}

\title{ Pure annihilation type $ D\to PP(V)$ decays in the perturbative QCD approach \thanks{ Supported by National
Science Foundation of China under the Grant No.11228512, 11235005 and
11075168}}

\author{
\quad ZOU Zhi-Tian
\quad LI Cheng
\quad L\"{U} Cai-Dian $^{1)}$ 
}
\maketitle

\address{
Institute  of  High  Energy  Physics  and  Theoretical  Physics Center for Science Facilities,
Chinese Academy of Sciences, Beijing 100049, People's Republic of China
}

\begin{abstract}
The annihilation type diagrams are difficult to calculate in any kind of models or method. Encouraged by the the successful calculation of pure annihilation type $B$ decays in the perturbative QCD factorization approach,   we calculate the pure annihilation type $D\to PP(V)$ decays in the perturbative QCD approach   based on the $k_T$ factorization. Although the expansion parameter $1/m_D$ is not very small, our leading order numerical results  agree with the existing experiment data for most channels. We expect the more accurate observation from experiments, which can help us learn about the dynamics of $D$ meson weak decays.
\end{abstract}

\begin{keyword}
D meson, perturbative QCD, annihilation
\end{keyword}

\begin{pacs}
13.20.Ft, 12.38.Bx, 14.40.Lb
\end{pacs}


\begin{multicols}{2}

\section{Introduction}

After decades of study,  the   $D$ meson decays are still a hot topic in both  theoretical side and experimental side, since they can provide useful information on flavor mixing, CP violation, strong interactions and even the new physics signal \cite{kaishi,cp1,cp2}. For example the recent observation of $D^0-\bar{D}^{0}$ mixing provides us a new platform to explore new physics via favor-changing neutral currents. By now, The CLEO-c and two B factories experiments have given many results about the $D$ decays. The BES-III experiment is expected to give more results. The accurate observation can help us understand the QCD dynamics and the $D$ meson weak decays. In recent years, many theoretical studies on the decays of $D$ meson have been done based on diagrammatic approach \cite{diagrammatic}, the final-state interaction effects \cite{yang1,yang2}, combination of factorization and pole model \cite{yu},  factorization assisted topological diagrammatic approach \cite{yucp}, and the perturbative QCD (PQCD) approach \cite{liyong}.

Most of the theoretical study show that  the annihilation type diagrams in hadronic D decays play a very important role \cite{diagrammatic,yu,yucp,liyong}.  For example in ref.\cite{diagrammatic}, the authors take the model-independent diagrammatic approach to study the two-body nonleptonic $D$ decays, with all topological amplitudes extracted from the experimental data. Their analysis indicates that the SU(3) breaking effect and the annihilation type contributions are important to explain the experimental data.The importance of annihilation diagram contribution  is also reflected from the large difference of  $D^0$ and $D^+$ lifetime. However, these   annihilation type  diagrams are usually very difficult to calculate, since factorization may not work here.   In ref.\cite{yu}, the authors use the pole model to give large annihilation diagram contributions.
It is worth of mentioning that the annihilation type diagrams can be perturbatively calculated without parametrization in the PQCD approach based on $k_T$ factorization \cite{annihilation1,annihilation2}. For these pure annihilation type $B$ decays, the predications in the PQCD approach have been confirmed by experiments later \cite{prd70034009,prd76074018,epjc28305,prd78014018}.

The factorization that is proved in the $1/m_b$ expansion, can be applied to the corresponding $D$ meson decays straightforwardly. However, the expansion is much poorer in D Decays than that in B decays due to smaller D meson mass. Anyway since there is no better method for the annihilation diagram calculation,   the pure annihilation type decays $D^0\to \bar{K}^{0}\phi$ were calculated in the PQCD approach \cite{liyong}, with a good agreement with the experimental result. In this work, we use the PQCD approach to   analyze the 10 modes of pure annihilation type $D\to PP(V)$ decays.   By keeping the intrinsic transverse momentum $k_T$ of valence quarks, the end point singularity, which will spoil the perturbative calculation, can be regulated by Sudakov form factor and threshold resummation.  Therefore, the PQCD approach can give converging results   with predictive power.

In standard model,  two body hadronic $D$ meson weak decays are dominated by the contributions from tree operators, since the contributions from the penguin operators are suppressed both by the small elements of the Cabibbo-Kobayashi-Maskawa (CKM) matrix and by the relatively small b quark mass in the $c-b-u$ penguin diagram. This is in contrast to the penguin amplitude in $B$ decays, which can profit from a larger CKM element and a much larger $t$ quark mass. Although the suppressed penguin diagram   contributions may be the main source of  the direct asymmetry \cite{cp1,cp2,yucp,cp3}, we ignore the penguin contributions in this work due to the small effect on the branching fractions.


\section{Formalism and Perturbative Calculation}

For the pure annihilation type $D\rightarrow PP(V)$ decays, at the quark level,   the dominant contributions are described by the effective Hamiltonian $H_{eff}$
\begin{eqnarray}
H_{eff}=\frac{G_{F}}{\sqrt{2}}\,V_{uq}V^{*}_{cq^{\prime}}\left[C_{1}(\mu)O_{1}(\mu)\,+\,C_{2}(\mu)O_{2}(\mu)\right],
\label{HH}
\end{eqnarray}
where $V_{cq^{\prime}}$ and $V_{uq}$ are the corresponding CKM matrix elements, with $q^{(\prime)}=d,s$
, and $C_{1,2}(\mu)$ are Wilson
coefficients at the renormalization scale $\mu$. $O_{1,2}(\mu)$ are
the four quark operators from tree diagrams
\begin{eqnarray}
O_{1}\,=\,(\bar{q}^{\prime}_{\alpha}c_{\beta})_{V-A}(\bar{u}_{\beta}q_{\alpha})_{V-A},
\;O_{2}\,=\,(\bar{q}^{\prime}_{\alpha}c_{\alpha})_{V-A}(\bar{u}_{\beta}q_{\beta})_{V-A},\nonumber
\end{eqnarray}
where $\alpha$ and $\beta$ are the color indices,
$(\bar{q}^{\prime}_{\alpha}c_{\beta})_{V-A}\,=\,\bar{q}^{\prime}_{\alpha}\gamma^{\mu}(1-\gamma^{5})c_{\beta}$.
Conventionally, the combination of Wilson coefficients can be defined as
\begin{eqnarray}
a_{1}=C_{2}+C_{1}/3,\;a_{2}=C_{1}+C_{2}/3.
\end{eqnarray}

In the hadronic matrix element calculation, the decay amplitude can be factorized into soft($\Phi$), hard(H), and
harder (C) dynamics characterized by different scales \cite{liyong,prd81014002},
\begin{eqnarray}
\mathcal
{A}\;\sim\;&&\int\,dx_{1}dx_{2}dx_{3}b_{1}db_{1}b_{2}db_{2}b_{3}db_{3}\nonumber\\
&&\times
Tr\left[C(t)\Phi_{D}(x_{1},b_{1})\Phi_{M_{2}}(x_{2})\Phi_{M_{3}}(x_{3})\right.\nonumber\\
&&\left.H(x_{i},b_{i},t)S_{t}(x_{i})e^{-S(t)}\right],
\end{eqnarray}
where  $b_{i}$ is the conjugate space coordinate of quark's transverse
momentum $k_{iT}$, $x_{i}$ is the momentum fractions of valence quarks, and $t$ is the largest energy scale in the hard part function
$H(x_{i},b_{i},t)$.  $C(t)$ are the Wilson coefficients with resummation of the large   QCD corrections of four quark operators. The large double logarithms $ln^{2}x_{i}$
 are summed by the threshold resummation to give a jet function $S_{t}(x_i)$
which smears the end-point singularities on $x_{i}$ \cite{prd66094010}. The Sudakov form
factor $e^{-S(t)}$ is from resummation of double logarithms, which suppresses the soft dynamics effectively and
the long distance contributions in the large $b$ region
\cite{prd57443,lvepjc23275}. Thus it makes the perturbative
calculation of the hard part $H$ reliable. The meson wave functions $\Phi_i$, are nonperturbative input parameters but universal for all decay modes.

\begin{center}
\vspace{-1.5cm}
\includegraphics[width=10cm]{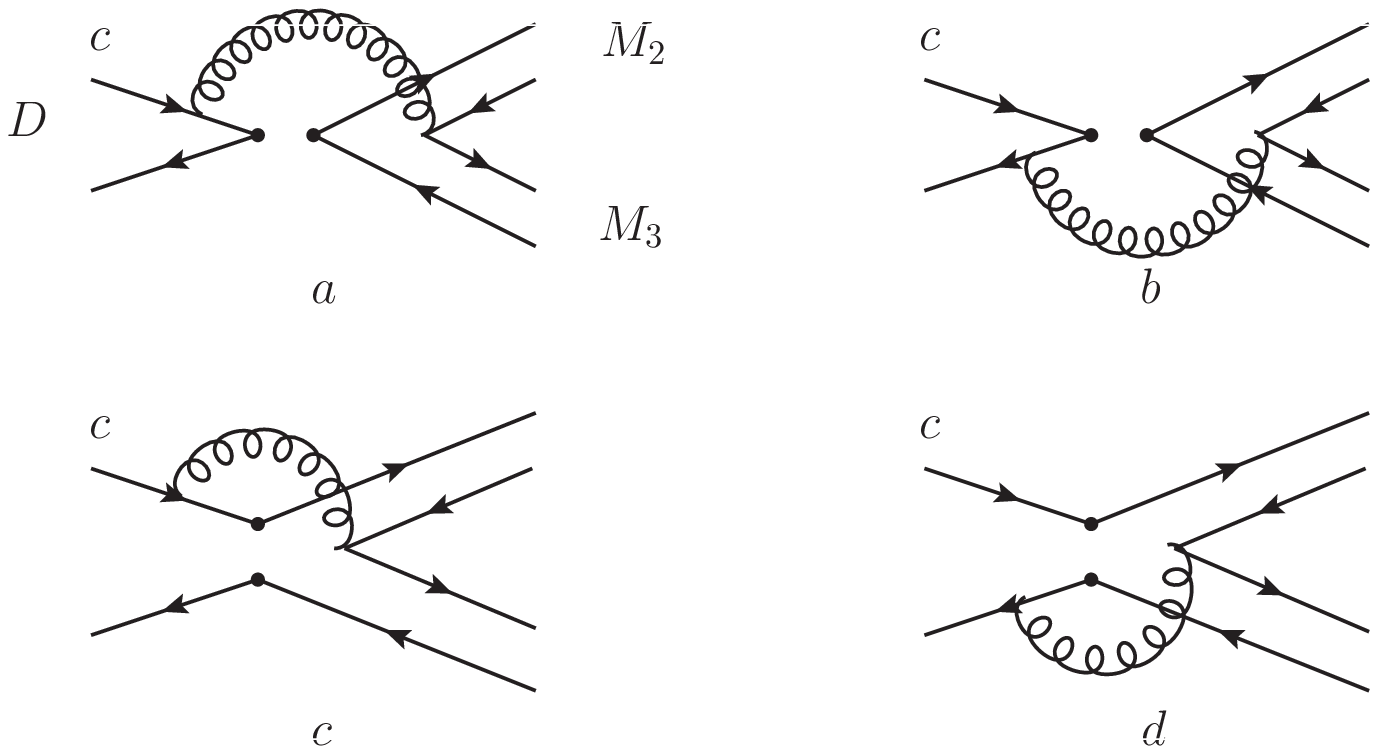}
\vspace{-9.4cm}
\figcaption{\label{fig}  The diagrams contributing to the pure annihilation type
$D\to PP(V)$ decays in PQCD }
\end{center}

The leading order Feynman diagrams of the considered decays are shown in Fig.1. For $D\to PP$ decays, the amplitude
from factorizable diagrams (a) and (b) in Fig.1 is
\begin{eqnarray}
\mathcal {A}_{af}&=&-8 C_{F}f_{D}\pi
m_{D}^{4}\int_{0}^{1}\,dx_{2}dx_{3}\int_{0}^{1/\Lambda}\,b_{2}db_{2}b_{3}db_{3}\nonumber\\
&&\times\left\{\left[2\phi_{M_2}^{P}(x_{2})r_{02}r_{03}(\phi_{M_3}^{P}(x_3)(x_3-2)-x_3\phi_{M_3}^{T}(x_3))\right.\right.\nonumber\\
&&\left.\left. +\phi^{A}_{M_{2}}(x_{2})\phi^{A}_{M_{3}}(x_{3})(x_3-1)\right] h_{af}(\alpha,\beta,b_{2},b_{3})E_{af}(t_{a})\right.\nonumber\\
&& +\left.\left[2\phi_{M_3}^{P}(x_3)r_{02}r_{03}(\phi_{M_2}^{T}(x_2)(x_2-1)+\phi_{M_2}^{P}(x_2)(x_2+1))\right.\right.\nonumber\\
&&\left.\left.+x_2\phi_{M_2}^{A}(x_2)\phi_{M_3}^{A}(x_3)\right] h_{af}(\alpha^{\prime},\beta,b_{3},b_{2})E_{af}(t_{b})\right\},
\label{af}
\end{eqnarray}
where, $C_{F}=4/3$ is the group factor of $SU(3)_{c}$, and $r_{02(03)}=m_{02(03)}/m_{D}$ with the chiral mass $m_{02(03)}$ of the pseudoscalar meson. The hard scale $t_{e,f}$ and the functions $E_{af}$ and $h_{af}$ can be given by
\begin{eqnarray}
t_{a}\,&=&\,\max\{\sqrt{(r_3^2+x_2(1-r_3^2))(1-r_2^2)(1-x_3)}m_{D},\nonumber\\
&&\sqrt{1-x_3(1-r_2^2)}m_D,1/b_{2},1/b_{3}\},\nonumber\\
t_{b}\,&=&\,\max\{\sqrt{(1-r_2^2)(r_3^2+x_2(1-r_3^2))}m_{D},\nonumber\\
&&1/b_{2},1/b_{3}\},
\label{taf}
\end{eqnarray}
\begin{eqnarray}
&&E_{af}(t)\,=\,\alpha_{s}(t)\cdot \exp[-S_{M_2}(t)-S_{M_3}(t)],
\end{eqnarray}
\begin{eqnarray}
h_{af}(\alpha,\beta,b_{2},b_{3})\,&=&\,(\frac{i\pi}{2})^{2}H_{0}^{(1)}\left(\beta b_{2}\right)S_{t}(x_{3})\nonumber\\
&&\left[\theta(b_{2}-b_{3})H_{0}^{(1)}\left(\alpha b_{2}\right)J_{0}\left(\alpha b_{3}\right)\right.\,+\nonumber\\
&&\left.\theta(b_{3}-b_{2})H_{0}^{(1)}\left(\alpha b_{3}\right)J_{0}\left(\alpha b_{2}\right)\right],
\end{eqnarray}
with $r_{2(3)}=m_{M_{2(3)}}/m_{D}$, $\alpha^2=(1-x_3(1-r_2^2))m_D^2$, $\beta^2=(r_3^2+x_2(1-r_3^2))(1-r_2^2)(1-x_3)m_D^2$ and $\alpha^{\prime2}=(r_3^2+x_2(1-r_3^2))(1-r_2^2)m_{D}^{2}$.

For the so called non-factorizable diagrams (c) and (d) in Fig.1, the decay amplitude is
\begin{eqnarray}
\mathcal {M}_{anf}&=&16\sqrt{\frac{2}{3}}C_{F}\pi
m_{D}^{4}\int_{0}^{1}\,dx_{1}dx_{2}dx_{3}\int_{0}^{1/\Lambda}\,b_{1}db_{1}b_{2}db_{2}\nonumber\\
&&\times \phi_{D}(x_{1},b_{1}) \left\{\left[\phi_{M_2}^{A}(x_{2})\phi_{M_3}^{A}(x_{3})(x_1+x_2)\right.\right.\nonumber\\
&&\left.\left.+r_{02}r_{03}\left(\phi_{M_2}^{P}(x_{2})\left(\phi_{M_3}^{P}(x_3)(x_1+x_2-x_3+3)\right.\right.\right.\right.\nonumber\\
&&\left.\left.\left.\left.+\phi_{M_3}^{T}(x_3)(1-x_1+x_2-x_3)\right)\right.\right.\right.\nonumber\\
&&\left.\left.\left.+\phi_{M_2}^{T}(x_2)\left(\phi_{M_3}^{P}(x_3)(x_1+x_2+x_3-1)\right.\right.\right.\right.\nonumber\\
&&\left.\left.\left.\left.+\phi_{M_3}^{T}(x_3)(x_3-x_1-x_2+1)\right)\right)\right]\right.\nonumber\\
&&\left.\cdot h_{anf1}(\alpha,\sqrt{|\beta_{1}^2|},b_{1},b_{2})E_{anf}(t_{c})\right.\nonumber\\
&&\left.+\left[\phi_{M_2}^{A}(x_2)\phi_{M_3}^{A}(x_3)(x_3-1)\right.\right.\nonumber\\
&&\left.\left.+r_{02}r_{03}\left(\phi_{M_2}^{P}(x_{2})\left(\phi_{M_3}^{P}(x_3)(x_1-x_2+x_3-1)\right.\right.\right.\right.\nonumber\\
&&\left.\left.\left.\left.+\phi_{M_3}^{T}(x_{3})(x_1-x_2-x_3+1)\right)\right.\right.\right.\nonumber\\
&&\left.\left.\left.+\phi_{M_2}^{T}(x_2)\left(\phi_{M_3}^{P}(x_3)(x_2+x_3-x_1-1)\right.\right.\right.\right.\nonumber\\
&&\left.\left.\left.\left.+\phi_{M_3}^{T}(x_3)(1-x_1+x_2-x_3)\right)\right)\right]\right.\nonumber\\
&&\cdot\left.h_{anf2}(\alpha,\sqrt{|\beta_{2}^2|},b_{1},b_{2})E_{anf}(t_{d})\right\},
\end{eqnarray}
with
\begin{eqnarray}
t_{g}&=&\max\{\sqrt{(r_3^2+x_2(1-r_3^2))(1-r_2^2)(1-x_3)}m_{D},\nonumber\\
&&\sqrt{1-[(1-r_3^2)(1-x_2)-x_1][r_2^2+x_3(1-r_2^2)]}m_{D},\nonumber\\
&&1/b_{1},1/b_{2}\},\nonumber\\
t_{h}&=&\max\{\sqrt{|(x_1-r_3^2-x_{2}(1-r_3^2))|(1-r_2^2)(1-x_3)}m_{D},\nonumber\\
&&\sqrt{(r_3^2+x_2(1-r_3^2))(1-r_2^2)(1-x_3)}m_{D},\nonumber\\
&&1/b_{1},1/b_{2}\},
\label{tanf}
\end{eqnarray}
\begin{eqnarray}
E_{anf}\,=\,\alpha_{s}(t)\cdot
\exp[-S_{D}(t)-S_{M_2}(t)-S_{M_3}(t)]\mid\,_{b_{2}=b_{3}},
\end{eqnarray}
\begin{eqnarray}
h_{anfj}&=&\frac{i\pi}{2}\left[\theta(b_{1}-b_{2})H_{0}^{(1)}\left(\alpha b_{1}\right)J_{0}\left(\alpha b_{2}\right)\right.\nonumber\\
&&\left.+\theta(b_{2}-b_{1})H_{0}^{(1)}\left(\alpha b_{2}\right)J_{0}\left(\alpha b_{1}\right)\right]\nonumber\\
&&\times \left\{\begin{array}{ll}
\frac{i\pi}{2}H_{0}^{(1)}\left(\sqrt{|\beta_{j}^2|} b_{1}\right),&
\beta_{j}^{2}<0,\\
K_{0}\left(\sqrt{|\beta_{j}^2|}b_{1}\right),& \beta_{j}^{2}>0,
\end{array}\right.
\end{eqnarray}
where $j=1,2$, $\beta_1^2=1-[(1-r_3^2)(1-x_2)-x_1][r_2^2+x_3(1-r_2^2)]m_{D}^2$, $\beta_2^2=(x_1-r_3^2-x_{2}(1-r_3^2))(1-r_2^2)(1-x_3)m_D^2$ , and $\alpha=\sqrt{(r_3^2+x_2(1-r_3^2))(1-r_2^2)(1-x_3)}m_D$.
The expressions of $S_{D}(t)$, $S_{M_2}(t)$, $S_{M_3}(t)$ and $S_{t}$ can be found in refs.\cite{prd66094010,lvepjc23275, epjc28515}.

For those $D\to PV$ decays, the decay amplitudes are
\begin{eqnarray}
\mathcal {A}_{af}^{PV}&=&8 C_{F}f_{D}\pi
m_{D}^{4}\int_{0}^{1}\,dx_{2}dx_{3}\int_{0}^{1/\Lambda}\,b_{2}db_{2}b_{3}db_{3}\nonumber\\
&&\times\left\{\left[2\phi_{M_2}^{P}(x_{2})r_{02}r_{V}(\phi_{V}^{s}(x_3)(x_3-2)-x_3\phi_{V}^{t}(x_3))\right.\right.\nonumber\\
&&\left.\left. +\phi^{A}_{M_{2}}(x_{2})\phi_{V}(x_{3})(r_V^2-1)(x_3-1)\right] \right.\nonumber\\ &&\left.\cdot h_{af}(\alpha,\beta,b_{2},b_{3})E_{af}(t_{a})\right.\nonumber\\
&& -\left.\left[-2\phi_{V}^{s}(x_3)r_{02}r_V(\phi_{M_2}^{T}(x_2)(x_2-1)+(x_2+1)\right.\right.\nonumber\\
&&\left.\left.\cdot\phi_{M_2}^{P}(x_2))+(x_2+(1-2x_2)r_V^2)\phi_{M_2}^{A}(x_2)\phi_{V}(x_3)\right]\right.\nonumber\\
&&\left.\cdot h_{af}(\alpha^{\prime},\beta,b_{3},b_{2})E_{af}(t_{b})\right\},
\label{af3}
\end{eqnarray}
\begin{eqnarray}
\mathcal {M}_{anf}^{PV}&=&16\sqrt{\frac{2}{3}}C_{F}\pi
m_{D}^{4}\int_{0}^{1}\,dx_{1}dx_{2}dx_{3}\int_{0}^{1/\Lambda}\,b_{1}db_{1}b_{2}db_{2}\nonumber\\
&&\times \phi_{D}(x_{1},b_{1}) \left\{\left[\phi_{M_2}^{A}(x_{2})\phi_{V}(x_{3})\right.\right.\nonumber\\
&&\left.\left.(x_1+x_2+(-x_1-2x_2+x_3+1)r_V^2))\right.\right.\nonumber\\
&&\left.\left.+r_{02}r_{V}\left(\phi_{M_2}^{T}(x_{2})\left(\phi_{V}^{s}(x_3)(1-x_1-x_2-x_3)\right.\right.\right.\right.\nonumber\\
&&\left.\left.\left.\left.+\phi_{V}^{t}(x_3)(x_1+x_2-x_3-1)\right)\right.\right.\right.\nonumber\\
&&\left.\left.\left.+\phi_{M_2}^{P}(x_2)\left(\phi_{V}^{t}(x_3)(x_1+x_2+x_3-1)\right.\right.\right.\right.\nonumber\\
&&\left.\left.\left.\left.-\phi_{V}^{s}(x_3)(x_1+x_2-x_3+3)\right)\right)\right]\right.\nonumber\\
&&\left.\cdot h_{anf1}(\alpha,\sqrt{|\beta_{1}^2|},b_{1},b_{2})E_{anf}(t_{c})\right.\nonumber\\
&&\left.-\left[\phi_{M_2}^{A}(x_2)\phi_{V}(x_3)(x_3-1)(2r_V^2-1)\right.\right.\nonumber\\
&&\left.\left.+r_{02}r_{V}\left(\phi_{M_2}^{P}(x_{2})\left(\phi_{V}^{s}(x_3)(x_1-x_2+x_3-1)\right.\right.\right.\right.\nonumber\\
&&\left.\left.\left.\left.+\phi_{V}^{t}(x_{3})(x_1-x_2-x_3+1)\right)\right.\right.\right.\nonumber\\
&&\left.\left.\left.+\phi_{M_2}^{T}(x_2)\left(\phi_{V}^{s}(x_3)(x_2+x_3-x_1-1)\right.\right.\right.\right.\nonumber\\
&&\left.\left.\left.\left.+\phi_{V}^{t}(x_3)(1-x_1+x_2-x_3)\right)\right)\right]\right.\nonumber\\
&&\cdot\left.h_{anf2}(\alpha,\sqrt{|\beta_{2}^2|},b_{1},b_{2})E_{anf}(t_{d})\right\},
\end{eqnarray}
with $r_V=r_3=m_{V}/m_{D}$.
For $D \to VP$ decays, the amplitudes are
\begin{eqnarray}
\mathcal {A}_{af}^{VP}&=&8 C_{F}f_{D}\pi
m_{D}^{4}\int_{0}^{1}\,dx_{2}dx_{3}\int_{0}^{1/\Lambda}\,b_{2}db_{2}b_{3}db_{3}\nonumber\\
&&\times\left\{\left[2\phi_{V}^{s}(x_{2})r_{03}r_{V}(\phi_{M_3}^{T}(x_3)x_3-\phi_{M_3}^{P}(x_3)(x_3-2))\right.\right.\nonumber\\
&&\left.\left. +\phi^{A}_{M_{3}}(x_{3})\phi_{V}(x_{2})((2x_3-1)r_V^2-x_3+1)\right]\right.\nonumber\\
&&\left.\cdot h_{af}(\alpha,\beta,b_{2},b_{3})E_{af}(t_{a})\right.\nonumber\\
&& -\left.\left[2\phi_{M_3}^{P}(x_3)r_{03}r_V(\phi_{V}^{t}(x_2)(x_2-1)\right.\right.\nonumber\\
&&\left.\left.+\phi_{V}^{s}(x_2)(x_2+1))-\phi_{M_3}^{A}(x_3)\phi_{V}(x_2)(r_V^2-1)x_2\right]\right.\nonumber\\ &&\left. \cdot h_{af}(\alpha^{\prime},\beta,b_{3},b_{2})E_{af}(t_{b})\right\},
\label{af2}
\end{eqnarray}
\begin{eqnarray}
\mathcal {M}_{anf}^{VP}&=&16\sqrt{\frac{2}{3}}C_{F}\pi
m_{D}^{4}\int_{0}^{1}\,dx_{1}dx_{2}dx_{3}\int_{0}^{1/\Lambda}\,b_{1}db_{1}b_{2}db_{2}\nonumber\\
&&\times \phi_{D}(x_{1},b_{1}) \left\{\left[\phi_{M_3}^{A}(x_{3})\phi_{V}(x_{2})\right.\right.\nonumber\\
&&\left.\left.(x_1+x_2+(-2x_1-2x_2+1)r_V^2))\right.\right.\nonumber\\
&&\left.\left.+r_{03}r_{V}\left(\phi_{M_3}^{T}(x_{3})\left(\phi_{V}^{s}(x_2)(1-x_1-x_2-x_3)\right.\right.\right.\right.\nonumber\\
&&\left.\left.\left.\left.+\phi_{V}^{t}(x_2)(1-x_1-x_2+x_3)\right)\right.\right.\right.\nonumber\\
&&\left.\left.\left.+\phi_{M_3}^{P}(x_3)\left(\phi_{V}^{t}(x_2)(x_1+x_2+x_3-1)\right.\right.\right.\right.\nonumber\\
&&\left.\left.\left.\left.+\phi_{V}^{s}(x_2)(x_1+x_2-x_3+3)\right)\right)\right]\right.\nonumber\\
&&\left.\cdot h_{anf1}(\alpha,\sqrt{|\beta_{1}^2|},b_{1},b_{2})E_{anf}(t_{c})\right.\nonumber\\
&&\left.-\left[\phi_{M_3}^{A}(x_3)\phi_{V}(x_2)\right.\right.\nonumber\\
&&\left.\left.\cdot (1-x_3+r_V^2(x_1-x_2+2x_3-2))\right.\right.\nonumber\\
&&\left.\left.+r_{03}r_{V}\left(\phi_{M_3}^{P}(x_{3})\left(\phi_{V}^{s}(x_2)(1-x_1+x_2-x_3)\right.\right.\right.\right.\nonumber\\
&&\left.\left.\left.\left.+\phi_{V}^{t}(x_{2})(x_1-x_2-x_3+1)\right)\right.\right.\right.\nonumber\\
&&\left.\left.\left.+\phi_{M_3}^{T}(x_3)\left(\phi_{V}^{s}(x_2)(x_2+x_3-x_1-1)\right.\right.\right.\right.\nonumber\\
&&\left.\left.\left.\left.+\phi_{V}^{t}(x_2)(x_1-x_2+x_3-1)\right)\right)\right]\right.\nonumber\\
&&\cdot\left.h_{anf2}(\alpha,\sqrt{|\beta_{2}^2|},b_{1},b_{2})E_{anf}(t_{d})\right\},
\end{eqnarray}
with $r_V=r_2=m_{V}/m_D$. The form of the wave functions of final state pseudoscalar mesons and vector mesons can be found in ref.\cite{prd76074018}, with the different Gegenbauer moments used in this work as
\begin{eqnarray}
&&a_{2\pi}^{A}=0.70,a_{4\pi}^{A}=0.45,a_{2\pi}^{P}=0.70,a_{4\pi}^{P}=0.36,\nonumber\\
&&a_{3\pi}^{T}=0.80,a_{1K}^{A}=0.60,a_{2K}^{A}=0.10,a_{2K}^{P}=0.5,\nonumber\\
&&a_{4K}^{P}=-0.2,a_{3K}^{T}=0.65,a_{2\rho}^{\|}=a_{2\omega}^{\|}=0.6,\nonumber\\
&&a_{2\phi}^{\|}=0.70,a_{1K^*}^{\|}=0.6,a_{2K^*}^{\|}=0.11.
\end{eqnarray}
Since the energy release in $D$ decays is smaller than that in  $B$ decays, our light meson wave functions have larger SU(3) breakings in $D$ decays. For the distribution amplitudes of $D/D_s$ meson, we take the same model as the $B$ meson \cite{prd76074018} with different hadronic parameter $\omega=0.35/0.5$ for $D/D_s$ meson.

With the functions obtained in the above, the amplitudes of
 these pure annihilation decay channels can be given by
\begin{eqnarray}
\mathcal {A}(D^{0}\rightarrow
K^{(*)0}\bar{K}^{(*)0})=\frac{G_{F}}{\sqrt{2}}\left\{V_{cd}^{*}V_{ud}\left[a_{2}\mathcal
{A}_{af}^{K^{(*)0}\bar{K}^{(*)0}}\right.\right.\nonumber\\
\left.\left.+C_{2}\mathcal {M}_{anf}^{K^{(*)0}\bar{K}^{(*)0}}\right]+V_{cs}^{*}V_{us}\left[a_{2}\mathcal
{A}_{af}^{\bar{K}^{(*)0}K^{(*)0}}\right.\right.\nonumber\\
\left.\left.+C_{2}\mathcal{M}_{anf}^{\bar{K}^{(*)0}K^{(*)0}}\right]\right\},
\label{1}
\end{eqnarray}
\begin{eqnarray}
\mathcal {A}(D^{0}\rightarrow
K^{0}\phi)=\frac{G_{F}}{\sqrt{2}}V_{cd}^{*}V_{us}[a_{2}\mathcal
{A}_{af}^{K\phi}+C_{2}\mathcal {M}_{anf}^{K\phi}],
\label{2}
\end{eqnarray}
\begin{eqnarray}
\mathcal {A}(D^{0}\rightarrow
\bar{K}^{0}\phi)=\frac{G_{F}}{\sqrt{2}}V_{cs}^{*}V_{ud}[a_{2}\mathcal
{A}_{af}^{\phi\bar{K}}+C_{2}\mathcal {M}_{anf}^{\phi\bar{K}}],
\label{8}
\end{eqnarray}
\begin{eqnarray}
\mathcal {A}(D^{+}\rightarrow
K^{+}\phi)=\frac{G_{F}}{\sqrt{2}}V_{cd}^{*}V_{us}[a_1\mathcal
{A}_{af}^{K\phi}+C_{1}\mathcal {M}_{anf}^{K\phi}],
\label{4}
\end{eqnarray}
\begin{eqnarray}
\mathcal {A}(D_s\rightarrow
\pi^{+}\pi^0)&=&\frac{G_{F}}{2}V_{cs}^{*}V_{ud}[a_{2}(\mathcal
{A}_{af}^{\pi^0\pi^+}-\mathcal
{A}_{af}^{\pi^+\pi^0})\nonumber\\
&&+C_{2}(\mathcal {M}_{anf}^{\pi^0\pi^+}-\mathcal {M}_{anf}^{\pi^+\pi^0})]\nonumber\\
&&\sim 0,
\label{5}
\end{eqnarray}
\begin{eqnarray}
\mathcal {A}(D_s\rightarrow
\pi^0\rho^+)&=&\frac{G_{F}}{2}V_{cs}^{*}V_{ud}[a_{2}(\mathcal
{A}_{af}^{\pi^0\rho^+}-\mathcal
{A}_{af}^{\rho^+\pi^0})\nonumber\\
&&+C_{2}(\mathcal {M}_{anf}^{\pi^0\rho^+}-\mathcal {M}_{anf}^{\rho^+\pi^0})],
\label{6}
\end{eqnarray}
\begin{eqnarray}
\mathcal {A}(D_s\rightarrow
\pi^+\rho^0(\omega))&=&\frac{G_{F}}{\sqrt{2}}V_{cs}^{*}V_{ud}[a_{2}(\mathcal
{A}_{af}^{\pi^+\rho^0(\omega)}\nonumber\\
&&\mp\mathcal{A}_{af}^{\rho^0(\omega)\pi^+})+C_{2}(\mathcal {M}_{anf}^{\pi^+\rho^0(\omega)}\nonumber\\
&&\mp\mathcal {M}_{anf}^{\rho^0(\omega)\pi^+})].
\label{7}
\end{eqnarray}

\section{Numerical Results and Discussions}

For numerical analysis, we use the following input parameters:
\begin{eqnarray}
&&f_{D/D_{s}}=0.23/0.257GeV,f_{K}=0.16GeV,f_{\pi}=0.13GeV,\nonumber\\
&&f_{\rho}^{(T)}=0.209(0.165)GeV, f_{K^{*}}^{(T)}=0.217(0.185)GeV,\nonumber\\ &&f_{\omega}^{(T)}=0.195(0.145)GeV,f_{\phi}^{(T)}=0.220(0.185)GeV,\nonumber\\
&&|V_{cd}|=0.2252\pm0.00065,|V_{ud}|=0.9742\pm0.0002,\nonumber\\
&&|V_{cs}|=0.97344\pm0.00016,|V_{us}|=0.2253\pm0.00065,\nonumber\\
&&
m_{0\pi}=1.4GeV, m_{0K}=1.6GeV, 
\Lambda_{QCD}^{f=3}=0.375GeV.
\end{eqnarray}

After numerical calculation, the branching ratios of these decays together with experimental measurements \cite{prd} are listed in Table~\ref{table}. We also list the results from diagrammatic approach \cite{diagrammatic} and pole model \cite{yu} for comparison.


The branching ratio obtained from the analytic formulas may be sensitive to many parameters especially those in
the meson wave function. The theoretical uncertainties in our calculations, shown  in Table~\ref{table}, are caused by the variation of (i)
the hadronic parameters, such as the shape parameters and the Gegenbauer moments in
wave functions of initial and final state
mesons; (ii) the unknown
next-to-leading order QCD corrections  and nonperturbative power corrections, characterized by the choice of the
$\Lambda_{QCD}\,=\,(0.375\,\pm\,0.05)$ GeV and the variations of the
factorization scales defined in eq.(\ref{taf}) and eq.(\ref{tanf}), respectively.

In hadronic $D$  decays, the SU(3) breaking effect is remarkable, which can be demonstrated by the decay channel $D^0\to K^{0}\bar{K}^{0}$, with large branching ratio from experimental measurement. There are two kinds contributions from the quark pair $d\bar{d}$ and $s\bar{s}$ produced through weak vertex. In SU(3) limit, the two contributions exactly cancel with each other due to the cancelation of the CKM matrix elements. Thus the diagrammatic approach \cite{diagrammatic} results in   zero branching ratio for this channel. Taking the SU(3) breaking effect in account, we give the result in agreement with the experimental data. For the decay $D^0\to \bar{K}^{0}\phi$, we reproduce the result of ref.\cite{liyong}, which agree well with the experimental data.
For $D_{s}^{+}\to \pi^{+}\pi^{0}$ decay, the branching ratio vanishes due to the exact cancelation of the contributions from $u\bar{u}$ and $d\bar{d}$ components. In fact, this decay is forbidden because the two pions can not form an $s$ wave isospin $1$ state due to the Bose-Einstein statistics. Any non-zero data for this decay may indicate the signal of new physics beyond the standard model.

\end{multicols}
\begin{center}
\tabcaption{ \label{table}  Branching ratios($10^{-3}$) for $D\to PP(V)$ decays together with experimental data \cite{prd}, the recent results from diagrammatic approach \cite{diagrammatic} and the predictions from pole model \cite{yu}.}
\small
\begin{tabular*}{170mm}{@{\extracolsep{\fill}}ccccc}
\toprule  decay modes & this work  & Br(diagrammatic)& Br(pole model) & Br(Exp) \\
\hline
$D^{0}\to K^{0}\bar{K}^{0}$ & $0.27^{+0.09}_{-0.08}$ & 0 & $0.3\pm0.1$ & $0.34\pm 0.08$ \\
$D_{s}\to \pi^{+}\pi^{0}$ & 0 & 0 & 0 & $<0.34$ \\
$D^0\to \bar{K}^{0}\phi$ & $8.55_{-3.41}^{+3.60}$ &$8.68\pm0.139$ & $0.8\pm0.2$ & $8.34\pm0.65$ \\
$D^0\to \bar{K}^{0}K^{*0}$ & $0.44_{-0.17}^{+0.20}$ &$0.29\pm0.22$ & $0.16\pm0.05$ & $<0.56$ \\
$D^0\to K^0\bar{K}^{*0}$ & $0.54_{-0.15}^{+0.20}$ &$0.29\pm0.22$ & $0.16\pm0.05$ & $<1.0$ \\
$D^0\to K^{0}\phi$ & $0.012_{-0.004}^{+0.004}$ &$0.006\pm0.005$ & $0.020\pm0.006$ &  \\
$D^+\to K^{+}\phi$ & $0.025_{-0.008}^{+0.012}$ & & $0.020\pm0.0020$ &  \\
$D_{s}^{+}\to \pi^{+}\rho^0$&$2.11_{-0.25}^{+0.87}$  &  & $4.0\pm4.0$& $0.2\pm0.12$ \\
$D_{s}^+ \to \pi^{+}\omega$ & $0.050_{-0.025}^{+0.029}$ & & 0 & $2.5\pm0.7$ \\
$D_s^+ \to \pi^{0}\rho^{+}$ & $2.11_{-0.24}^{+0.87}$ & & $4.0\pm4.0$ &  \\
\bottomrule
\end{tabular*}%
\end{center}
\begin{multicols}{2}


For $D_s^+\to \pi^{+}\rho^{0}$ decay, the branching ratio is larger than the experimental result, while for $D_s^+\to \pi^+\omega$, it is much smaller than the experimental result. The reason  is   that the minus sign of $\rho^{0}$ relative to $\omega$ is compensated by the asymmetric space wave function of the two final states which are in the P-wave state. One possible solution  is  the soft final-state interactions  as discussed in ref.\cite{diagrammatic}. In general, the soft final state interaction should be important in $D$ meson decays, because there are  many resonance states near the $D$ meson mass, which may give severe pollution to $D$ decays calculation.   We expect more accurate measurements from experiments such as LHCb and BESS-III, which can help us understand better the dynamics of $D$ meson decays.

\section{Summary}

In this work, we calculate the branching ratio of the 10 pure annihilation type $D_{(s)}\to PP(V)$ decays
in the perturbative QCD factorization approach without considering soft final states interactions. For most channels, our results   agree well with the experimental data. The SU(3) breaking effect is  found to be remarkable, which can be indicated by the large branching ratio of $D^{0}\to K^0\bar{K}^{0}$ decay. We hope that the super B factories and BES-III
can provide more accurate measurements for these decays, which will help us learn about the  QCD dynamics in $D$ meson decays and the annihilation mechanism.

\acknowledgments{We are very grateful to Yu Xin and
Yu Fu-Sheng for helpful discussions. 
}

\end{multicols}
\begin{multicols}{2}

\end{multicols}

\clearpage


\begin{thebibliography}{90}

\vspace{3mm}

\bibitem{kaishi} Arfuso M, Neadows B, Petrov A. A, Ann. Rev. Nucl. Part. Sci, 2008, \textbf{58}, 249-291
\bibitem{cp1}Pirtskhalava D, Uttayarat P, Phys. Lett B, 2012, \textbf{712}, 81¨C86
\bibitem{cp2}Bhattacharya B, Gronau M, Rosner J. L, Phys. Rev. D, 2012 \textbf{85}, 054014
\bibitem{diagrammatic} Cheng H. Y, Chiang C. W, Phys. Rev. D, 2010, \textbf{81}, 074021
\bibitem{yang1} Ablikim M, Du D. S, Yang M. Z, High Energy Phys. Nucl. Phys, 2003, \textbf{27},759-766
\bibitem{yang2} Li J. W, Yang M. Z, Du D. S, High Energy Phys. Nucl. Phys, 2003, \textbf{27}, 665-672
\bibitem{yu} Yu F. S, Wang X. X, L\"{u} C. D, Phys. Rev. D, 2011, \textbf{84}, 074019
\bibitem{yucp} Li H. N, L\"{u} C. D, Yu F. S, Phys. Rev. D, 2012, \textbf{86}, 036012
\bibitem{liyong} Du D. S, Li Y, L\"{u} C. D, Chin. Phys. Lett, 2006, \textbf{23}, 2038-2041
\bibitem{annihilation1}
 L\"{u} C. D, Ukai K, Eur. Phys. J. C, 2003, \textbf{28}, 305
[arXiv:hep-ph/0210206].
\bibitem{annihilation2}
Li Y, L\"{u} C. D, J. Phys. G, 2003, \textbf{29}, 2115; High Energy Phys. \& Nucl. Phys, 2003,
\textbf{27}, 1062
\bibitem{prd70034009}
Li Y, L\"{u} C. D, Xiao Z. J, Yu X. Q, Phys. Rev. D, 2004, \textbf{70}, 034009
\bibitem{prd76074018}
Ali A \textit{et} \textit{al}., Phys. Rev. D, 2007, \textbf{76}, 074018
\bibitem{epjc28305}
 L\"{u} C. D, Ukai K, Eur. Phys. J. C, 2003, \textbf{28}, 305
\bibitem{prd78014018}
Li R. H, L\"{u} C. D, Zou H, Phys. Rev. D, 2008, \textbf{78} 014018
\bibitem{cp3} Brod J, Kagan A. L, Zupan J, Phys.Rev. D, 2012 \textbf{86}, 014023

\bibitem{prd81014002}
Liu X, Xiao Z. J, L\"{u} C. D, Phys. Rev. D, 2010, \textbf{81}, 014002
\bibitem{prd66094010}
Li H. N, Phys. Rev. D, 2002, \textbf{66}, 094010
\bibitem{prd57443}
Li H. N, Tseng B, Phys. Rev. D, 1998, \textbf{57}, 443
\bibitem{lvepjc23275}
L\"{u} C. D, Yang M. Z, Eur. Phys. J. C, 2002, \textbf{23}, 275-287
\bibitem{epjc28515}
 L\"{u} C. D, Yang M Z, Eur. Phys. J. C, 2003, \textbf{28}, 515
\bibitem{prd}
Beringer J \textit{et} \textit{al}. (Partical Data Group), Phys. Rev. D, 2012, \textbf{86}, 010001


\end{thebibliography}
\end{document}